\documentclass[aps,prl,twocolumn,preprintnumbers,groupedaddress,nofootinbib]{revtex4}

\usepackage{amssymb}
\usepackage{epsfig} 
\usepackage{amsmath,color} 
\usepackage{graphicx} 
\usepackage{ulem} 
\usepackage{tabularx}
\usepackage{slashed}

\usepackage{tikz}
\usetikzlibrary{positioning,arrows}
\usetikzlibrary{decorations.pathmorphing}
\usetikzlibrary{decorations.markings}

\newcommand{\beq}{\begin{equation}}
\newcommand{\eeq}{\end{equation}}
\newcommand{\bea}{\begin{eqnarray}}
\newcommand{\eea}{\end{eqnarray}}
\newcommand{\ba}{\begin{array}}
\newcommand{\ea}{\end{array}}
\newcommand{\bi}{\begin{itemize}}
\newcommand{\ei}{\end{itemize}}
\newcommand{\bn}{\begin{enumerate}}
\newcommand{\en}{\end{enumerate}}
\newcommand{\bc}{\begin{center}}
\newcommand{\ec}{\end{center}}
\renewcommand{\l}{\left}
\renewcommand{\r}{\right}

\newcommand{\eq}[1]{Eq.~(\ref{#1})}
\newcommand{\eqs}[2]{Eqs.~(\ref{#1}) and (\ref{#2})}

\newcommand{\MeV}{\mathinner{\mathrm{MeV}}}

\begin{document}

\preprint{FTUV-16-08-13}
\preprint{IFIC-16-61}

\title{Flavor versus mass eigenstates in neutrino asymmetries: implications for cosmology.}


\author{Gabriela Barenboim$^1$}
\email[]{Gabriela.Barenboim@uv.es}
\author{William H.\ Kinney$^2$} 
\email[]{whkinney@buffalo.edu}
\author{Wan-Il Park$^{1,3}$}
\email[]{wipark@jbnu.ac.kr}
\affiliation{$^1$
Departament de F\'isica Te\`orica and IFIC, Universitat de Val\`encia-CSIC, E-46100, Burjassot, Spain}

\affiliation{$^2$ Dept. of Physics, University at Buffalo, 239 Fronczak Hall, Buffalo, NY 14260-1500}

\affiliation{$^3$ Division of Science Education and Institute of Fusion Science, Chonbuk National University, Jeonju 54896, Republic of Korea}


\date{\today}

\begin{abstract}
We show that, if they exist, lepton number asymmetries ($L_\alpha$) of neutrino flavors should be distinguished from the ones ($L_i$) of mass eigenstates, since Big Bang Nucleosynthesis (BBN)  bounds on the  flavor eigenstates cannot be directly applied to the mass eigenstates. Similarly, Cosmic Microwave Background (CMB) constraints on mass eigenstates do not directly constrain flavor asymmetries.
Due to the difference of mass and flavor eigenstates, the cosmological constraint on the asymmetries of neutrino flavors can be much stronger than conventional expectation, but not uniquely determined unless at least the asymmetry of the heaviest neutrino is well constrained.
Cosmological constraint on $L_i$ for a specific case is presented as an illustration.
\end{abstract}


\maketitle


\section{Introduction}

A large lepton number asymmetry of neutrinos is an intriguing possibility with respect to its capability of resolving several non-trivial issues of cosmology (see for example \cite{Bajc:1997ky,Liu:1993am,MarchRussell:1999ig}), but has been known to be constrained tightly by BigBang nucleosynthesis (BBN) \cite{Mangano:2010ei,Mangano:2011ip}.
Interestingly, in a recent paper \cite{largeL} it has been shown that, even if BBN constrains the lepton number asymmetry of the electron-neutrino very tightly such as $L_e \lesssim \mathcal{O}(10^{-3})$, much larger muon- and tau-neutrino asymmetries of $\mathcal{O}(0.1-1)$ are still allowed as long as the total lepton number asymmetry is sizeable.
Such large asymmetries are expected to be constrained mainly by cosmic microwave background (CMB) via the extra neutrino species $\Delta N_{\rm eff}$ \cite{Ade:2015xua}. 

If asymmetric neutrinos have a thermal distribution, their contribution to $\Delta N_{\rm eff}$ is expressed as 
\beq \label{dNeff}
\Delta N_{\rm eff} = \frac{15}{7} \sum_\alpha \l( \frac{\xi_\alpha}{\pi} \r)^2 \l[ 2 + \l( \frac{\xi_\alpha}{\pi} \r)^2 \r]
\eeq
where $\xi_\alpha \equiv \mu_\alpha/T$ is the neutrino degeneracy parameter. 
Conventionally, the summation in \eq{dNeff} has been done with neutrino flavors ($\nu_{e,\mu,\tau}$ in case of only three active neutrinos).
An implicit assumption here is that the extra radiation energy coming from asymmetric neutrinos are solely from flavor-eigenstates.
However, due to neutrino flavor oscillations \cite{Fukuda:1998mi,Ahmad:2001an,An:2012eh,Ahn:2012nd}, the equilibrium density matrix is not diagonal in the flavour basis (as one naively expects, flavor eigenstates not being asymptotic states of the Hamiltonian) and their description in terms of only diagonal components (a more or less hidden assumption when assuming thermal distribution for flavors) cannot capture all the contributions to the extra radiation energy density \cite{Starkman:1999zy}.
On the other hand, well after their decoupling from a thermal bath, free-streaming neutrinos should be described as incoherent mass-eigenstates only.
Hence, the appropriate estimation of $\Delta N_{\rm eff}$ should be done exclusively with neutrino mass-eigenstates instead of flavor-eigenstates in \eq{dNeff}.

In this letter, we argue that the equilibrium lepton number asymmetry matrix reached by the  BBN epoch is diagonal in the mass eigenstate basis and  related to the one in flavor eigenstate basis simply by the Pontecorvo-Maki-Nakagawa-Sakata (PMNS) matrix, and show that the lepton number asymmetries of the mass eigenstates are different from those of flavors.
A numerical demonstration is provided.
Also, we discuss implications of a lepton number asymmetry on cosmological data such as CMB+SNIa.

\section{Lepton number asymmetries of neutrino flavor vs. mass eigenstates}

\begin{figure*}[ht]
\begin{center}
\includegraphics[width=0.47\textwidth]{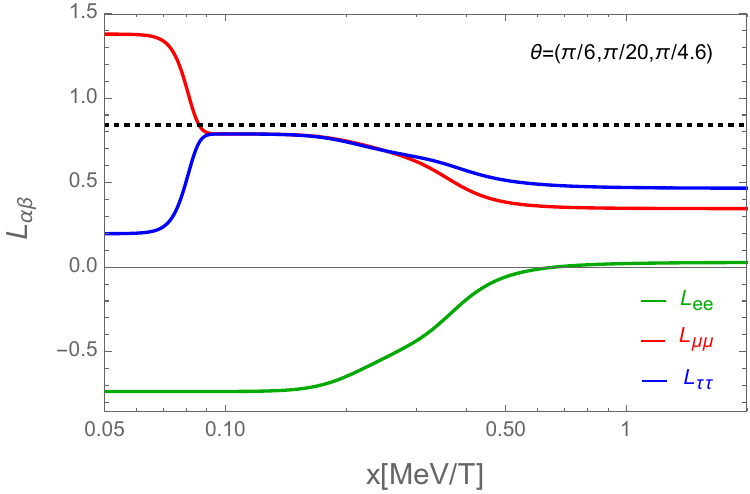}
\includegraphics[width=0.47\textwidth]{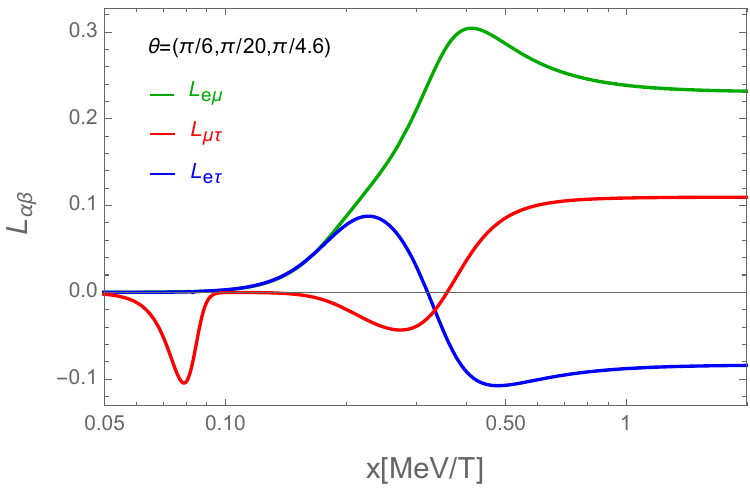}
\caption{Evolutions of $\mathbf{L}_{\rm f}$ for $\theta=(\theta_{12},\theta_{13},\theta_{23})$ with $\theta_{ij}$ being the mixing angles in PMNS matrix, and $(\xi_e,\xi_\mu,\xi_\tau) = (-1.0,1.6,0.3)$.
\textit{Left/Right}: Diagonal/off-diagonal entries.
}
\label{fig:Lf}
\end{center}
\end{figure*}
%
\begin{figure*}[ht]
\begin{center}
\includegraphics[width=0.47\textwidth]{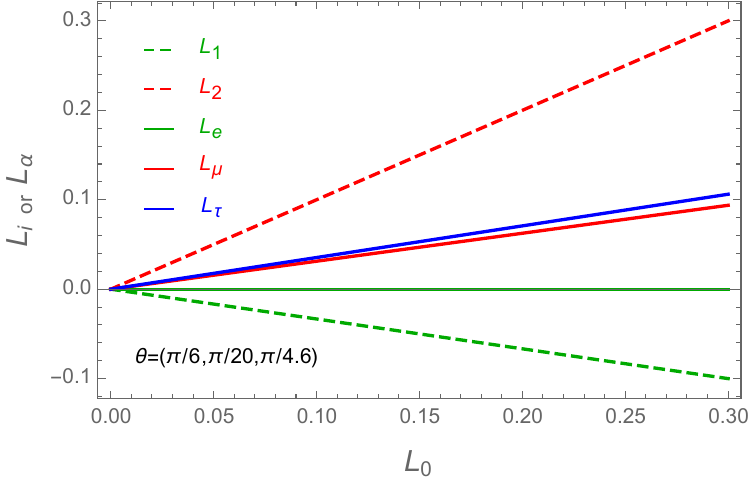}
\includegraphics[width=0.47\textwidth]{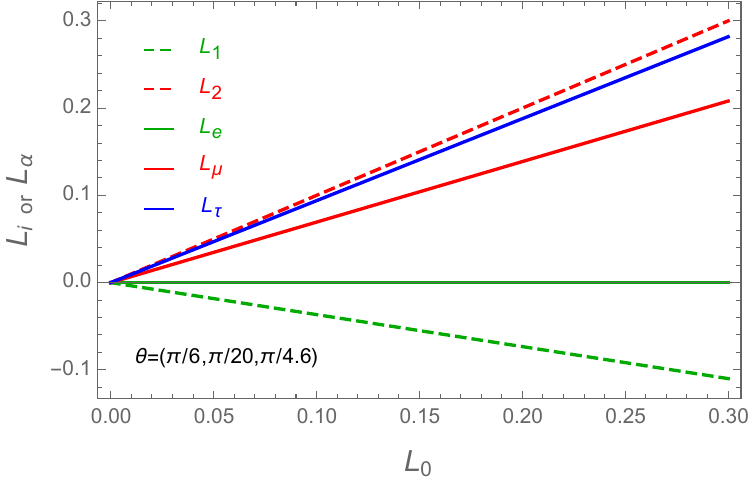}
\caption{Comparisons of lepton number asymmetries of both mass-eigenstates ($L_i ; \ i=1,2,3$) and flavor-eigenstates ($L_\alpha ; \ \alpha = e,\mu,\tau$) for $\theta =(\theta_{12},\theta_{13},\theta_{23})$.
Solid/dashed lines are the asymmetries of flavor/mass eigenstates.
Left and right panels are showing two examples of $\mathbf{L}_{\rm m}$ leading to $L_e \approx 0$ satisfying BBN constraint.
\textit{Left}: $\mathbf{L}_{\rm m} = {\rm diag}(L_1,L_2,L_3)=(-t_{12}^2 L_0, L_0, 0)$.
\textit{Right}: $\mathbf{L}_{\rm m} = {\rm diag}(-(t_{12}^2+t_{13}^2/c_{12}^2) L_0, L_0, L_0)$.
}
\label{fig:L-CMB}
\end{center}
\end{figure*}
%

The lepton number asymmetries of neutrinos in flavor basis can be defined as a matrix such as
\beq
\mathbf{L}_{\rm f} = \frac{\rho-\bar{\rho}}{n_\gamma}
\eeq
where $\rho/\bar{\rho}$ and $n_\gamma$ are the (mode-integrated) density matrices of neutrinos/anti-neutrinos and the photon number density.
In the very early universe, it is natural to assume that neutrinos are in interaction eigenstates (i.e., flavor eigenstates), since their kinematic phases are very small and collisional interactions to thermal bath are large enough to block flavor oscillations.
Hence, if it were generated at very high energy, $\mathbf{L}_{\rm f}$ is likely to be diagonal and to remain constant. 
While oscillations  are blocked, individual flavor lepton numbers are conserved. 
However, due to  the fact that neutrinos  are not  massless  and mix (according to the values of  the mixing parameters and mass differences measured by a variety of experiments \cite{Agashe:2014kda}), as the temperature of the radiation dominated universe drops below around $T \sim 15 \MeV$, flavor oscillations become active.
$\mathbf{L}_{\rm f}$ starts evolving at this epoch, and settles down to an equilibrium state finally at $T \sim 2-5 \MeV$ before BBN starts  \cite{Lunardini:2000fy,Dolgov:2002ab,Wong:2002fa,Abazajian:2002qx,Mangano:2010ei,Castorina:2012md}, depending on the neutrino mass hierarchy.
Here, we consider the case of the normal mass hierarchy with zero CP-violating phase.

Once it reaches its final equilibrium value, $\mathbf{L}_{\rm f}$ becomes time-independent.
The shape of $\mathbf{L}_{\rm f}$ at the final equilibrium is determined by various effects including vacuum oscillations, MSW-like effect coming from charged lepton backgrounds, neutrino self-interactions,  and collisional damping.
So, it is difficult to predict analytically, and in practice is only accesible via numerical methods.  
However, all these effects except vacuum oscillations are active during particular windows in temperature and 
eventually disappear.
Hence, the final shape of $\mathbf{L}_{\rm f}$ should be determined by vacuum oscillation parameters only.
Note that the flavor states mixed by vacuum oscillation parameters are nothing but mass-eigenstates in flavor-basis. 
Therefore, the statistical equilibrium state of $\mathbf{L}_{\rm f}$ should be that of mass-eigenstates expressed in the flavor-basis.

Since in vacuum mass- and flavor-eigenstates are related to each other by the PMNS matrix, $U_{\rm PMNS}$ \cite{Maki:1962mu,Pontecorvo:1957cp}, our argument implies that for a diagonalization matrix $D$, the matrix $\mathbf{L}_{\rm m}$ of asymmetries in mass basis is given by
\beq \label{L-mass}
\mathbf{L}_{\rm m} = D^{-1} \mathbf{L}_{\rm f} D = U_{\rm PMNS}^{-1} \mathbf{L}_{\rm f} U_{\rm PMNS}
\eeq
implying 
\beq \label{D-matrix}
D=U_{\rm PMNS}
\eeq
On general grounds, at late times we do not expect $\mathbf{L}_{\rm f}$ to be diagonal. The operator responsible for the evolution of the density matrix is not diagonal, so that a diagonal density matrix will not be the asymptotic solution of those equations unless it is proportional to the identity matrix.  
Hence, generically the asymmetries of neutrino mass eigenstates differ from those of flavor, and this fact should be taken into account when observational constraints on lepton number asymmetries are considered.

In order to verify our argument, we solved numerically the quantum kinetic equations of neutrino/anti-neutrino density matrices \cite{Sigl:1992fn,Pantaleone:1992eq} in a simplified way as done in Ref.~\cite{largeL} in which the dynamics of a typical mode mimicking the collective behavior of all modes was analyzed.
An example is shown in Fig.~\ref{fig:Lf} where one finds the evolutions of $L_{\alpha \beta}$, the (real) entries of $\mathbf{L}_{\rm f}$ for the neutrino's normal mass hierarchy with \cite{Agashe:2014kda}
\bea
\Delta m_{21}^2 &=& 7.53 \times 10^{-5} {\rm eV}^2
\\
\Delta m_{31}^2 &\simeq& \Delta m_{32}^2 = 2.67 \times 10^{-3} {\rm eV}^2
\eea
and the mixing angle $\theta_{ij}$ shown in the figure \cite{Agashe:2014kda,NOvA}.
As shown in the right panel of the figure, the off-diagonal entries of $\mathbf{L}_{\rm f}$ do not disappear, making $\mathbf{L}_{\rm m}$ be different from $\mathbf{L}_{\rm f}$.
Also, we found that the numerical simulation reproduces the relation \eq{D-matrix} quite precisely within errors of $\mathcal{O}(0.1) \%$ even at $x=1$.

The differences between diagonal entries of $\mathbf{L}_{\rm f}$ and $\mathbf{L}_{\rm m}$ can be seen by expressing the former in terms of the latter.
At first, $L_e$ is given by 
\beq \label{Le-from-Li}
L_e = c_{13}^2 \l( c_{12}^2 L_1 + s_{12}^2 L_2 \r) + s_{13}^2 L_3,
\eeq
where $c_{ij}/s_{ij}/t_{ij} = \cos \theta_{ij}/ \sin \theta_{ij}/\tan \theta_{ij}$ with $\theta_{ij}$ being the mixing angle in PMNS matrix.
Since BBN requires $|L_e| \lesssim \mathcal{O}(10^{-3})$, we may set $L_e=0$ for an illustration when $|L_e| \lll |L_i|$ in \eq{Le-from-Li}.
In this case, $L_\mu$ and $L_\tau$ are given by
\bea \label{Lmu}
L_\mu &=& c_{23} \l[ (1-t_{12}^2) c_{23} -  2 s_{13} s_{23} t_{12} \r] L_2
\nonumber \\
&+& \l[ (1-t_{13}^2) s_{23}^2 - t_{12} t_{13}^2 c_{23} (2 s_{13} s_{23} + t_{12} c_{23}) \r] L_3, \phantom{1ex}
\\ \label{Ltau}
L_\tau &=& s_{23} \l[ (1-t_{12}^2) s_{23} + 2 s_{13} c_{23} t_{12} \r] L_2
\nonumber \\
&+& \l[ (1-t_{13}^2) c_{23}^2 + t_{12} t_{13}^2 s_{23} (2 s_{13} c_{23} - t_{12} s_{23}) \r] L_3. \phantom{1ex}
\eea
From \eqs{Lmu}{Ltau} with measured values of mixing angles \cite{Agashe:2014kda}, we find that $L_\mu \sim L_\tau$ for $|L_3| \lesssim |L_2|$, as shown in Fig.~\ref{fig:L-CMB}.
One may think that it is also possible to have $|L_{\mu,\tau}| \ll |L_{2,3}|$ if $L_2 \sim - L_3$.
However, our numerical testing showed that generically ${\bf Max}[\{ |L_{\alpha \beta}|_{\alpha \neq \beta} \}] \lesssim {\bf Max}[\{ |L_{\alpha \alpha}| \}]$.
Hence, on general grounds one expects to have 
\beq
\mathcal{O}(0.1) \lesssim {\bf Max}[\{ |L_i| \}]/{\bf Max}[\{ |L_{\alpha \alpha}| \}] \lesssim \mathcal{O}(1),
\eeq 
showing that it is critical to know at least two of $L_i$s in order to constrain $L_\mu$ and $L_\tau$.

\section{Cosmological constraints} 

A large lepton number asymmetry in one or more neutrino species creates an extra radiation density in the universe relative to the standard contributions of photons and CP-symmetric active neutrinos, a form of so-called ``dark radiation''. Extra relativistic degrees of freedom in cosmology have attracted considerable recent attention as a way to resolve the apparent discrepancy in measurement of the Hubble parameter from CMB data and type-Ia supernovae \cite{Ade:2015xua,Riess:2016jrr,Bernal:2016gxb,DiValentino:2016hlg,Archidiacono:2016kkh,Tram:2016rcw,Aghanim:2016sns,Ko:2016uft,Zhao:2016ecj}. In this section, we investigate the possibility that a primordial lepton asymmetry may provide a dark radiation density which can reconcile CMB and SNIa values for the Hubble parameter.   

We consider two basic cases. The first is an eight-parameter $\Lambda$CDM+$\xi$ cosmology without contribution from primordial tensor fluctuations, with parameters
\begin{itemize}
\item{Baryon density $\Omega_{\rm b} h^2$.}
\item{Dark matter density $\Omega_{\rm C} h^2$.}
\item{Angular scale of acoustic horizon $\theta$.}
\item{Reionization optical depth $\tau$.}
\item{Helium fraction $Y_P$.}
\item{Power spectrum normalization $A_s$.}
\item{Scalar spectral index $n_{\rm S}$.}
\item{Lepton asymmetry $\xi$.}
\end{itemize}
In the second case, motivated by models of early-universe inflation, we include the tensor/scalar ratio $r$ as a ninth parameter to the fit. $H_0$ is a derived parameter.
We assume a normal mass hierarchy for neutrinos, with one massive neutrino with mass $m_\nu = 0.06\ {\rm eV}$.
Since the BBN constraint on $L_e$ should be satisfied, we are not free to choose $|L_i| \gg |L_e|$ in an arbitrary way, but constrained to satisfy approximately
\beq
c_{12}^2 L_1 + s_{12}^2 L_2 + t_{13}^2 L_3 = \frac{L_e}{c_{13}^2} \approx 0,
\eeq
from \eq{Le-from-Li}.
As the simplest possibility, we may set $L_3=0$ leading to $L_1 \approx - t_{12}^2 L_2$.
Then, for thermal distributions of two light mass eigenstates \footnote{Strictly speaking \eq{eq:CMBNeff} is only valid for massless neutrinos. With non-zero masses, the thermal distribution and energy density of neutrinos/antineutrinos are modified (see, for example, Ref. \cite{Lesgourgues:1999wu}).
We have not taken into account such modifications in this work.
However, as long as the masses of two light mass eigenstates are much smaller than their momentum around the epoch of CMB decoupling, \eq{eq:CMBNeff} is a good enough approximation.},  
\begin{eqnarray}
\label{eq:CMBNeff}
\Delta N_{\rm eff} 
&&= \frac{15}{7} \sum_{i=1,2} \l( \frac{\xi_i}{\pi} \r)^2 \l[ 2 + \l( \frac{\xi_i}{\pi} \r)^2 \r]
\nonumber \\
&&\approx \frac{15}{7} \l( \frac{\xi_2}{\pi} \r)^2 \times
\nonumber \\
&& \l\{ (1+t_{12}^4) 2 + \l[ 1 + (4+t_{12}^4) t_{12}^4 \r] \l( \frac{\xi_2}{\pi} \r)^2 \r\}, \phantom{ex}
 \end{eqnarray}
where $\xi_i$s are degeneracy parameters of each mass eigenstate, and $|\xi_i| \lesssim 1$ and $t_{12}^2 \ll 1$ were assumed. (See also Refs. \cite{Schwarz:2012yw,Stuke:2011wz,Schwarz:2009ii} for a discussion of joint constraints on $N_{eff}$ and $Y_P$.)

Strictly speaking, the late-time free-streaming neutrino mass-eigenstates are not in thermal distribution since they are linear combinations of thermal distributions of flavor-eigenstates.
Hence, $\xi_i$s in \eq{eq:CMBNeff} should be understood as effective degeneracy parameters.
The error in $\Delta N_{\rm eff}$ depends on the initial configuration of the lepton number asymmetries in flavor-basis, but it is expected to be of or small than $\mathcal{O}(10)$\% for $|\xi_i| \lesssim 1$.
We constrain the parameter space with: (a) Planck 2015 TT/TE/EE+lowTEB temperature and polarization data \cite{Adam:2015rua,Ade:2015xua}, and the Bicep/Keck 2014 combined polarization data \cite{Ade:2015fwj}, and (b) CMB data combined with the Riess, {\it et al.} supernova data \cite{Riess:2016jrr}. The allowed contours are calculated numerically using a Markov Chain Monte Carlo method with the cosmomc software package \cite{Lewis:2002ah}, using the CAMB Boltzmann code modified according to Eq. (\ref{eq:CMBNeff}).\footnote{The data sets themselves contain multiple internal parameters, which we do not list here.} Curvature $\Omega_{\rm k}$ is set to zero, and the Dark Energy equation of state is fixed at $w = -1$.   For these constraints, we run 8 parallel chains with Metropolis-Hastings sampling, and use a convergence criterion of the Gelman and Ruben $R$ parameter of $R - 1 < 0.05$.

\subsection{Case 1: $\Lambda$CDM+$\xi$}

\begin{figure*}[h!t]
\begin{center}
\includegraphics[width=0.8\textwidth]{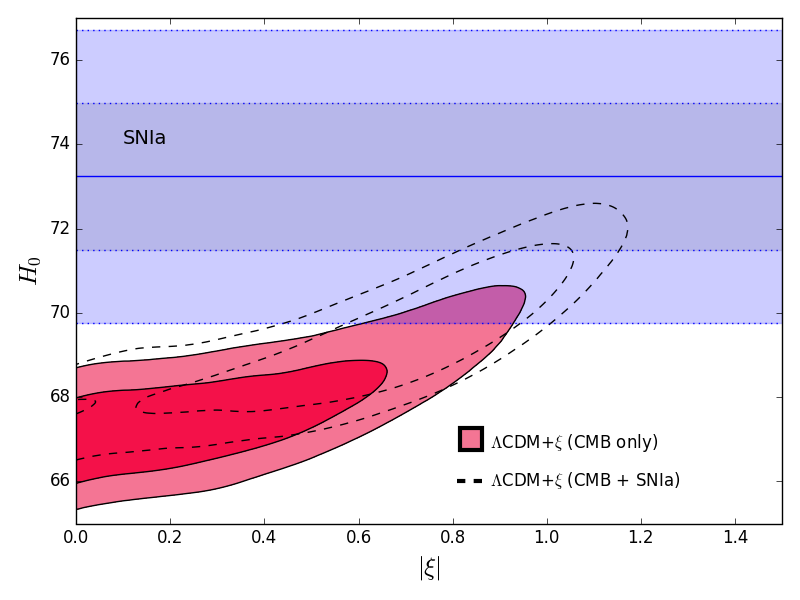}
\caption{Constraints on $H_0$ and $\xi$ for the eight-parameter $\Lambda$CDM+$\xi$ case. Filled contours show the 68\% (dark red) and 95\% (light red) constraints from Planck+BICEP/Keck alone. Dashed contours show the corresponding constraints with the addition of the Riess {\it et al.} supernova data. The constraint on $H_0$ from the supernova data alone, $H_0 = 73.24 \pm 1.74$ \cite{Riess:2016jrr} is shown by the filled regions, with $1\sigma$ limits in lavender, and $2\sigma$ limits in grey.}
\label{fig:CMB1}
\end{center}
\end{figure*}
%
\begin{figure*}[h!t]
\begin{center}
\includegraphics[width=0.49\textwidth]{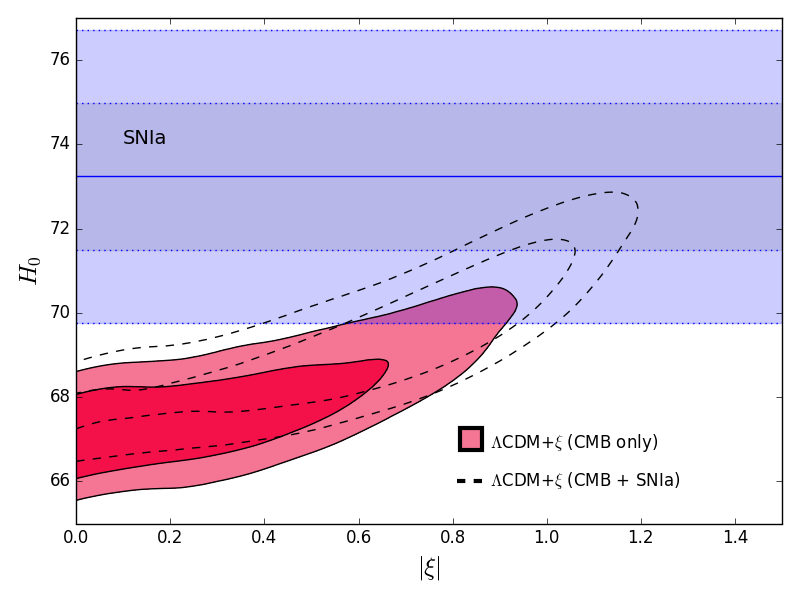}
\includegraphics[width=0.49\textwidth]{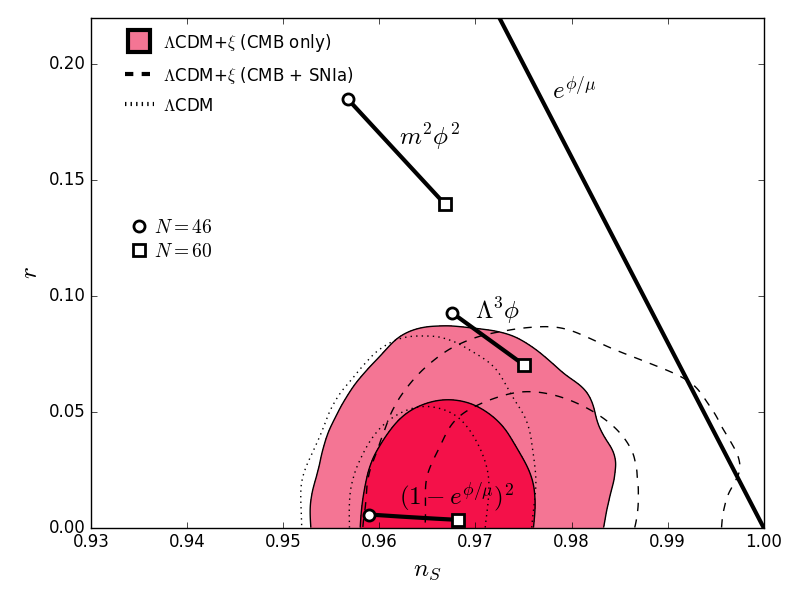}
\caption{CMB constraints on lepton number asymmetries for the nine-parameter model including tensor perturbations. 
Contours are 68\% and 95\% uncertainties from CMB-only (red-shaded regions), and CMB+supernovae (dashed lines).
\textit{Left}: Constraint as a function of $\xi$ and $H_0$. The filled region is the Riess {\it et al.} constraint on $H_0$ from supernovae.
\textit{Right}: Constraint on the spectral index $n_S$ and tensor-to-scalar ratio $r$, plotted with the predictions of representative choices of inflationary scalar-field potential. Dotted contours for $\Lambda$CDM+$r$, with fixed $\xi = 0$. This can be compared to Fig. 1 of Tram, {\it et al.} \cite{Tram:2016rcw}.
}
\label{fig:CMB2}
\end{center}
\end{figure*}
%
Figure \ref{fig:CMB1} shows constraints on $H_0$ and $\xi$ for the case of the eight-parameter $\Lambda$CDM+$\xi$ fit. We plot constraints from Planck+BICEP/Keck only (filled contours), and Planck+BICEP/Keck+Riess {\it et al.} (dashed contours). The CMB data alone show no evidence for nonzero neutrino chemical potential, with a 95\%-confidence upper bound of $|\xi| < 0.77$ (corresponding to $\Delta N_{\rm eff} \lesssim 0.30$ or $(L_1,L_2) \lesssim (-0.19,0.56)$ for $\theta_{12}=\pi/6$), with $H_0 = 67.71 \pm 0.95$. For combined CMB and supernova data, there is weak evidence for a nonzero chemical potential, with $|\xi| = 0.63 \pm 0.27$ (corresponds to $\Delta N_{\rm eff} \approx 0.20^{+0.21}_{-0.14}$ or $(L_1,L_2) \approx (-0.15^{+0.07}_{-0.07},0.45^{+0.22}_{-0.20})$ for $\theta_{12}=\pi/6$) at 68\% confidence, with $H_0 = 69.25 \pm 1.18$. 
The combined CMB+supernova data, however, should be interpreted with caution: as the filled contours illustrate, the CMB data and supernova data taken separately are barely compatible, with only a small overlap in the 95\% confidence regions, even when dark radiation from a neutrino asymmetry is included as a parameter. Combining two fundamentally incompatible data sets in a Bayesian analysis is likely to give a biased fit, which is reflected in the best-fit values for the two cases, with the best-fit to CMB alone having $-\ln({\mathcal L}) = 6794.38$, while the best-fit for the the combined CMB+supernova data is measurably worse, with $-\ln({\mathcal L}) = 6798.21$. For the CMB data alone, including lepton asymmetry, the 95\%-confidence upper bound on the Hubble parameter is $H_0 < 69.7$. This can be compared with a 95\%-confidence {\it lower} bound from Type-Ia supernovae of $H_0 > 69.8$. Other parameters are consistent with their best-fit $\Lambda$CDM values. We therefore conclude, contrary to existing claims in the literature \cite{Caramete:2013bua,Wyman:2013lza,Aubourg:2014yra,Bernal:2016gxb,Zhao:2016ecj}, that inclusion of dark radiation does not provide a consistent mechanism for reconciling the discrepancy between CMB and supernova data. Furthermore, there is no evidence for a nonzero lepton asymmetry from current data.

\subsection{Case 2: $\Lambda$CDM+$\xi$+$r$: constraints on inflation}

Figure \ref{fig:CMB2} shows parameter constraints on the nine-parameter case, with tensor perturbations included, consistent with generic expectations from inflation. Constraints in the $H_0$, $\xi$ parameter space are extremely similar to the case of no tensors, which is reasonable considering the upper bound of $r < 0.07$ obtained from Planck+BICEP/Keck data \cite{Kinney:2016qyl}. In this case we obtain a 95\%-confidence upper bound on the lepton asymmetry of $|\xi| < 0.77$, and $|\xi| = 0.63 \pm 0.29$ for Planck+BICEP/Keck+SNIa at 68\%-confidence. The best-fit to CMB alone is $-\ln({\mathcal L}) = 6793.52$, and CMB+SNIa is $-\ln({\mathcal L}) = 6798.14$, nearly identical to the no-tensor case. As in the no-tensor case, we conclude that here is no evidence for dark radiation from a lepton asymmetry. Constraints on inflationary potentials are shown in the right-hand panel of Fig. \ref{fig:CMB2}, which can be compared to Fig. 1 of Tram, {it et al.} \cite{Tram:2016rcw}. Our constraints here are considerably tighter. The difference is that here we include the BICEP/Keck polarization data, which results in a considerably stronger constraint on the parameter space than that provided by Planck alone. Of particular note, our constraint rules out power-law inflation, with $V\left(\phi\right) \propto e^{\phi / \mu}$, even in the presence of dark radiation, which is allowed by the constraints of Tram, {\it et al.} Ref. \cite{DiValentino:2016ucb} reaches a similar conclusion based on constraints from Planck on $\sigma_8$ and the reionization optical depth $\tau_{\rm reio}$.

\section{Conclusions}

In this letter, we argued that, when lepton number asymmetries of neutrinos in flavor basis are mixed among themselves due to neutrino oscillation  in the early universe before BBN, the eventual asymmetries after reaching the final equilibrium of flavor-mixings are well described in the basis of mass eigenstates, which are related to flavor eigenstates by the Pontecorvo-Maki-Nakagawa-Sakata (PMNS) matrix.
That is, the matrices of lepton number asymmetries in mass- and flavor-basis ($\mathbf{L}_{\rm m}$ and $\mathbf{L}_{\rm f}$, repectively) are related as
\beq
\mathbf{L}_{\rm m} = U_{\rm PMNS}^{-1} \mathbf{L}_{\rm f} U_{\rm PMNS},
\eeq
where $U_{\rm PMNS}$ is the PMNS matrix, and $\mathbf{L}_{\rm m}$ appears to be diagonal.
We demonstrated this argument by a numerical simulation, and showed analytically that the asymmetries of mass-eigenstates can be even larger than those of flavor-eigenstates.

Conventionally, the constraint on the lepton number asymmetries of neutrino flavors has been associated with neutrino flavor-eigenstates, counting their contributions to the extra radiation energy density $\Delta N_{\rm eff}$.
However, our finding showed that, when neutrino flavor-eigenstates have large lepton number asymmetries at temperatures well above $\mathcal{O}(10) \MeV$, neutrino flavor-mixings cause not only re-distribution of asymmetries among flavor-eigenstates but also sizable amounts of asymmetries of flavor-mixed states.
Hence, an appropriate estimation of $\Delta N_{\rm eff}$ should take into account the contributions from flavor-mixed states too.
Such an estimation can be performed in either flavor basis or mass basis, but mass basis provides a simpler way since the asymmetries are diagonal in the basis.
The resulting $\Delta N_{\rm eff}$ can be larger than the one estimated with flavor-eigenstates only.
This implies that the constraint on the lepton number asymmetries of neutrino flavor-eigenstates become stronger than conventional expectation (or the asymmetries of neutrino flavor-eigenstates are more constrained than those of mass-eigenstates). 

As shown in Ref.~\cite{largeL} and in this work, in principle $\Delta N_{\rm eff}$ can be of $\mathcal{O}(0.1-1)$ just from asymmetric neutrinos without resorting to an unknown ``dark radiation''.
Such a large $\Delta N_{\rm eff}$ has been considered in literature as a possible solution to the discrepancy of the measured expansion rate $H_0$ in CMB and SNIa data.
In analyses of cosmological data, typically, if $\Delta N_{\rm eff}$ is from asymmetric neutrinos, the neutrino degeneracy parameters have been taken in an arbitrary way without distinguishing mass- and flavor-eigenstates, although implicitly the lepton number asymmetry ($L_e$) of electron-neutrinos must be assumed to be small to satisfy BBN constraint.
We showed that this approach is inconsistent unless the lepton number asymmetries ($L_i$) of mass-eigenstates which are relevant for CMB data for example are constrained to satisfy 
\beq
L_e = c_{12}^2 L_1 + s_{12}^2 L_2 + t_{13}^2 L_3 \approx 0,
\eeq
for $|L_e| \lll |L_i|$.
Also, analyzing cosmological data (CMB only or CMB+SNIa), we found that CMB data alone show no evidence for nonzero neutrino lepton number asymmetries, with 95\% CL upper bound of $|\xi| \leq 0.77$ at $95$\% CL as the degeneracy parameter of the dominant mass eigenstate.
For combined CMB and SNIa data, there is weak evidence for  nonzero lepton number asymmetries, with $|\xi| = 0.63 \pm 0.27$ at 68\% CL, but the fit becomes worse relative to the case of CMB data alone.
So, even if large lepton number asymmetries may fit the data, it does not look preferred.

As the final remark, because of the degeneracy of $\mathbf{L}_{\rm m}$ for a given $\Delta N_{\rm eff}$, the bound on $\Delta N_{\rm eff}$ can not be uniquely interpreted in terms of the asymmetries of neutrino flavors (specifically $L_\mu$ and $L_\tau$ of muon- and tau-neutrinos), unless the impact on small scale power spectrum is sensitive enough to distinguish at least the contribution of the heaviest neutrinos.

\section{Acknowledgements}
Authors thank Selene Kinney for her helpful conversations.
GB acknowledges support from the MEC and FEDER (EC) Grants SEV-2014-0398 and FPA2014-54459 and the Generalitat Valenciana under grant PROMETEOII/2013/017. This project has received funding from the European Union's Horizon 2020
research and innovation programme under the Marie Sklodowska-Curie grant
Elusives ITN agreement No 674896  and InvisiblesPlus RISE, agreement No 690575. 
WHK is supported by the U.S. National Science Foundation under grant NSF-PHY-1417317. 
WHK and WIP thanks the University of Valencia, where part of this work was completed, for hospitality and support. 
This work was performed in part at the University at Buffalo Center for Computational Research. 
This work was also supported by ``Research Base Construction Fund Support Program'' funded by Chonbuk National University in 2017, and by Basic Science Research Program through the National Research Foundation of Korea (NRF) funded by the Ministry of Education (No. 2017R1D1A1B06035959).
We thank the referee for insightful comments.


\end{document}